# Electromagnetic response of the surface states of a topological insulator nanowire embedded within a resonator

Shimon Arie Haver[1], Eran Ginossar[2], Sebastian E. de Graaf[3] & Eytan Grosfeld[1]✉

Exploring the interplay between topological phases and photons opens new avenues for investigating novel quantum states. Here we show that superconducting resonators can serve as sensitive probes for properties of topological insulator nanowires (TINWs) embedded within them. By combining a static, controllable magnetic flux threading the TINW with an additional oscillating electromagnetic field applied perpendicularly, we show that orbital resonances can be generated and are reflected in periodic changes of the Q-factor of the resonator as a function of the flux. This response probes the confinement of the two-dimensional Dirac orbitals on the surface of the TINW, revealing their density of states and specific transition rules, as well as their dependence on the applied flux. Our approach represents a promising cross-disciplinary strategy for probing topological solid state materials using state-of-the-art photonic cavities, which would avoid the need for attaching contacts, thereby enabling access to electronic properties closer to the pristine topological states.

[1] Department of Physics, Ben-Gurion University of the Negev, Be'er-Sheva 84105, Israel. [2] Advanced Technology Institute and Department of Physics, University of Surrey, Guildford GU2 7XH, UK. [3] National Physical Laboratory, Teddington TW11 0LW, UK. ✉email: grosfeld@bgu.ac.il

Three-dimensional topological insulators (TIs) have generated intensive interest since their theoretical prediction[1–3] and experimental discovery[4,5]. They admit electronic surface states with spin-momentum locking[6] that, through the proximity effect, can give rise to remarkable electronic phases, including the anomalous quantum hall effect[7] and topological superconductivity[8]. Confined devices such as TI nanowires (TINWs) offer opportunities for probing the surface states and their magnetoelectric response in scenarios in which finite-size quantization plays an important role[9–12].

Coupling topological states to photons could lead to new testing grounds for topology[13–19] and to the generation of novel quantum states of matter. Superconducting resonators enhance the local electric field and facilitate high-resolution microwave probing in a highly coherent setup, with significant recent advances include designs with extremely high Q-factors[20] and tunable resonator frequency combined with magnetic field resilience[21]. Such high-Q resonators are becoming superb detectors of internal degrees of freedom generating losses. Embedding solid-state materials within resonators could open new frontiers for probing electronic properties, using relatively non-invasive techniques derived from the interaction with the photons, circumventing the need to attach contacts; while the possibility of applying static magnetic fields allows for an important control knob. These advances in TINWs and resonator fabrication offer a unique opportunity for detecting the structure of the surface state excitation spectrum and electromagnetic (EM) transition rules for confined TINWs, which remains an outstanding challenge.

Here we show that planar superconducting resonators can act as sensitive detectors for dissipation effects in the TINWs, while an applied static magnetic flux controls their properties, see Fig. 1. By calculating the EM response of the photon-induced oscillatory fields, we show that the Q-factor of the resonator gets reduced via the resonant photon absorption by the TINW, thus measuring this channel for photon loss. The relative ratio of the electric and magnetic field components can be controlled via the position of the TINW in the resonator. In principle, the dominant contribution is generated by the electric field. While screening can reduce this time-dependent density response, its effect strongly diminishes due to the presence of disorder and the strong dielectric environment. For the magnetic field component, a measurement of the Q-factor accesses an analog of the electron-spin-resonance (ESR) response of the TINW. However, we find that the gyromagnetic response is generated mainly by the orbital motion of the electrons around the circumference of the TINW, generating an "electron-orbital-resonance". This gives rise to a strongly enhanced effective gyromagnetic ratio, two orders of magnitude larger than the expected value from ESR alone. We study the effect of disorder and find that the EM absorption persists if the circumference of the TINW is smaller than the scattering length, which is consistent with experimentally available wires[9]. Our study indicates that the EM response could be measurable using state-of-the-art resonators.

## Results

**Model for surface state quantum electrodynamics.** We consider a TINW embedded within a single mode photonic resonator with a static magnetic flux $\Phi$ threading the wire. The full Hamiltonian for the system is $H = H_{nw} + H_{ph} + H_{int}$, where $H_{nw}$ describes the TINW, $H_{ph}$ describes the photons, and $H_{int}$ the light-matter interactions.

The Hamiltonian governing the surface states of the TINW in cylindrical geometry is[22–24]

$$H_{nw} = \iint dz d\varphi \, \Psi^\dagger(z,\varphi) \mathcal{H}_{nw} \Psi(z,\varphi), \quad (1)$$

with $\mathcal{H}_{nw} = \hbar\Omega\left[\sigma_z\left(\hat{\ell} + \frac{1}{2} - \eta\right) - \sigma_y \hat{k} R\right]$ the Hamiltonian density; $\sigma_j$ ($j = x, y, z$) the Pauli matrices; $\hat{k} = -i\partial_z$ and $\hat{\ell} = -i\partial_\varphi$ proportional to the momentum and orbital angular momentum along the TINW, respectively; $\eta = \Phi/\Phi_0$ the dimensionless flux (with $\Phi_0 = h/e$); and, $\Omega = v_F/R$ the angular frequency associated with the electron motion around the circumference of the TINW with $R$ its radius and $v_F$ its Fermi velocity. The electron field operator is $\Psi(z,\varphi) = \frac{1}{\sqrt{2\pi L}} \sum_{k\ell s} e^{ikz} e^{i\varphi\ell} \Psi_{k\ell s} c_{k\ell s}$, where $c_{k\ell s}$ is the electron annihilation operator associated with the orbital ($\ell \in \mathbb{Z}$) and linear momentum ($k$) mode along the $z$-direction, with $L$ the length of the TINW and $s = \pm$ the band index. The wavefunctions $\Psi_{k\ell s} \equiv U_{k\ell}^\dagger |s\rangle$ with $U_{k\ell} = \exp\left[\frac{i}{2}\sigma_x \text{atan2}\left(\ell + \frac{1}{2} - \eta, kR\right)\right]$ (atan2($x, y$) being the four-quadrant inverse tangent) have associated energies $\epsilon_{k\ell s} = s\hbar\Omega\sqrt{\left(\ell + \frac{1}{2} - \eta\right)^2 + (kR)^2}$.

The Hamiltonian of the single mode resonator is,

$$H_{ph} = \hbar\omega\left(a^\dagger a + \frac{1}{2}\right), \quad (2)$$

where $\omega$ is the frequency of the resonator, and $a^\dagger$ ($a$) is the creation (annihilation) of a photon in the resonator. By placing the TINW near a node or an anti-node of the resonator, the nanowire experiences magnetic or electric fields, or, in other points, a combination of these two fields. An orbital coupling is generated by the replacement $\mathbf{p} \to \mathbf{\pi} = \mathbf{p} - e\mathbf{A}$ (neglecting the Zeeman term), leading altogether to $\mathcal{H}_{int} = \mathbf{J} \cdot \mathbf{A} + \varrho\phi$ where $\mathbf{A}$ is the vector potential, $\mathbf{J} = ev_F\left(-\sigma_z e_\varphi + \sigma_y e_z\right)$ the current ($e_j$ are the unit basis vectors), $\varrho$ the charge density and $e$ the electron charge. We take the magnetic field along the $x$-direction $\mathbf{B} = B_0(a + a^\dagger)e_x$ and the electric field along the $y$-direction $\mathbf{E} = -iE_0(a - a^\dagger)e_y$ (see Fig. 1). The vector potential can be chosen to be $\mathbf{A} = B_0 y e_z(a + a^\dagger)$ and the scalar potential $\phi = -\mathbf{E} \cdot \mathbf{r}$, leading to the coupling

$$H_{int} = -i(a - a^\dagger)g_E(\xi_E + \xi_E^\dagger) + (a + a^\dagger)g_B(\xi_B + \xi_B^\dagger), \quad (3)$$

with $\xi_E$ ($\xi_B$) representing the electric (magnetic) dipole coupling of the TINW, associated with $\Delta k = 0$ and $\Delta \ell = 1$: $\xi_{E\setminus B} = \sum_{k\ell ss'} \Xi_{k\ell ss'}^{E\setminus B} c_{k,\ell,s}^\dagger c_{k,\ell-1,s'}$ where $(\Xi_{k\ell ss'}^E, \Xi_{k\ell ss'}^B) = \Psi_{k,\ell,s}^\dagger(\sigma_0, \sigma_y)\Psi_{k,\ell-1,s'}$, and $(g_E, g_B) = eR/2(E_0, v_F B_0)$. The amplitudes $E_0$ and $B_0$ are related to the zero-point fields via $E_0 = E_{zp} \cos(\omega z_0/c)$ and $B_0 = B_{zp} \sin(\omega z_0/c)$ where $z_0$ is the position of the TINW within the resonator and $c$ is the speed of light.

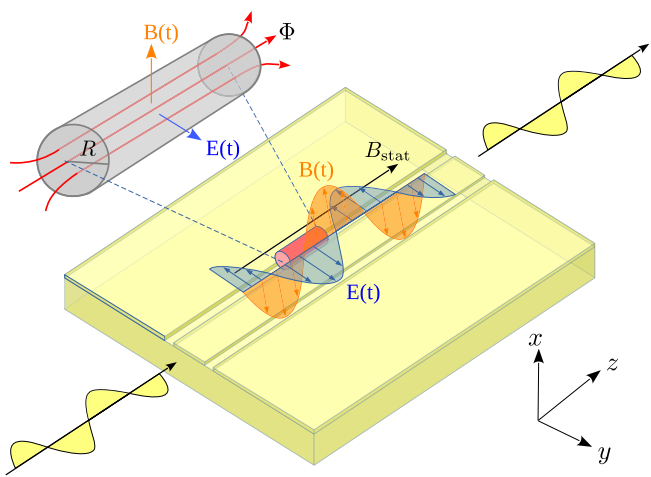

**Fig. 1 Schematic of a topological insulator nanowire in a resonator.** A topological insulator nanowire (TINW) of radius $R$ is embedded in a planar resonator with an oscillating electric ($E(t)$) and magnetic ($B(t)$) fields applied perpendicularly to the TINW. A static flux $\Phi = B_{stat}\pi R^2$ is threaded through the TINW, generated by an external static magnetic field, $B_{stat}$.

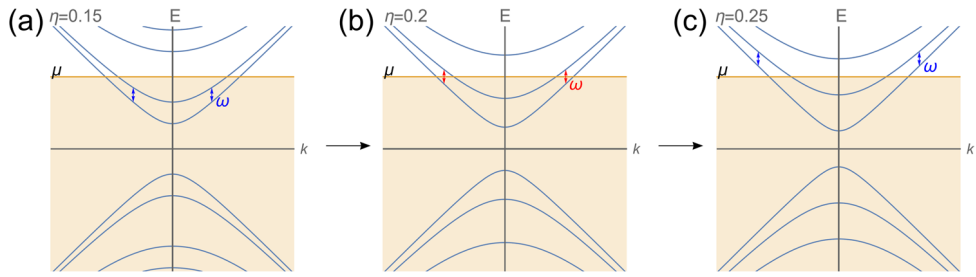

**Fig. 2 Subband spectrum of the topological insulator nanowire as function of the threaded flux.** Changes in the subband spectrum for three values of the dimensionless flux $\eta$ through the topological insulator nanowire (TINW): **a** $\eta = 0.15$; **b** $\eta = 0.2$; and, **c** $\eta = 0.25$. Here the chemical potential is taken to be $\mu = \hbar\Omega$, with $\hbar$ the reduced Planck constant and $\Omega$ the angular frequency associated with the electron motion around the circumference of the TINW. Resonant transitions with angular frequency $\omega = 0.2\Omega$ are depicted by blue and red arrows. Photon-induced transitions between subbands in the clean, noninteracting limit, occur if the condition for a resonance coincides with a transition from a full subband to an empty subband (**b**; red arrow); changing the flux $\eta$ shifts the energy levels and moves the resonance condition away from the chemical potential (**a**, **c**; blue arrows).

Finally, in the presence of disorder, we replace the Hamiltonian for the TINW by $\mathcal{H}_{nw} + V(z, \varphi)$, where $V(z, \varphi)$ is the disorder potential, and assume that the disorder is drawn from a Gaussian distribution, satisfying $\overline{V(z, \varphi)} = 0$, $\overline{V(z, \varphi) V(z', \varphi')} = \frac{\nu}{R}\delta(z'-z)\delta(\varphi'-\varphi)$, where the overline indicates averaging over the disorder realizations, and $\nu$ is a measure of the disorder strength with dimensions of energy squared times area.

**Electromagnetic response of the nanowire.** The magnetic and electric fields can generate orbital and spin responses within the TINW, controlled by the generation of magnetic and electric dipole moments. These are particularly significant when they lead to transitions between the subbands (see Fig. 2). In the following we will calculate the response functions for both cases and extract their dependence on the parameters of the model and the external flux.

The excitation of the TINW results from absorption of photons from the resonator. We assume that the electrons are further coupled to a phonon bath that takes away the excess energy so the system remains in the linear response regime. Hence to calculate the dissipation induced by the TINW in the resonator, we need to calculate the rate of absorption of photons. To perform this we employ the theoretical construct of adiabatically turning on the interaction. This allows for the extraction of the linewidth of the resonator via the regression hypothesis. To second order in the coupling to the photons we get the Kubo type formula for $n_{ph} = \langle a^\dagger a \rangle$, (see the Methods for additional information regarding the derivation)

$$\frac{dn_{ph}}{dt} = \lim_{\zeta \to 0} \frac{1}{\hbar^2} \int_{-\infty}^{t} dt' e^{\zeta t'} \sum_\alpha g_\alpha^2 \Big\langle \Big[ \big(a_I(t) - \iota_\alpha a_I^\dagger(t)\big)\big(\xi_{I\alpha}(t) + \xi_{I\alpha}^\dagger(t)\big), \big(a_I(t') + \iota_\alpha a_I^\dagger(t')\big)\big(\xi_{I\alpha}(t') + \xi_{I\alpha}^\dagger(t')\big) \Big] \Big\rangle, \quad (4)$$

where $\alpha = B, E$ and $\iota_B = 1$, $\iota_E = -1$. We next write, approximately,

$$\frac{dn_{ph}}{dt} \simeq -n_{ph} \sum_\alpha \gamma^\alpha(\omega), \quad (5)$$

where in the last transition we neglected the spontaneous emission from the TINW. This last equation defines the TINW excitation rate, $\gamma^\alpha(\omega)$, denoting the contributions due to the magnetic and electric fields.

The two absorption rates can be rewritten using the retarded correlation function $\chi_{\alpha\beta}^R$,

$$\gamma^\alpha(\omega) = \frac{i}{\hbar} g_\alpha^2 \sum_{j\in\pm} \left[ j\chi_{\alpha\alpha}^R(j\omega) - \text{h.c.} \right], \quad (6)$$

where in time representation $\chi^R(t)$ is given by[25],

$$\chi_{\alpha\beta}^R(t-t') = -\frac{i}{\hbar} \Theta(t-t') \Big\langle \Big[ \xi_\alpha(t), \xi_\beta^\dagger(t') \Big] \Big\rangle. \quad (7)$$

Writing the correlation function using the retarded and advanced Green's functions $G_{k\ell s}^{R/A}(\epsilon) = \left(\epsilon - \epsilon_{k\ell s} \pm i\frac{\hbar}{2\tau(\epsilon)}\right)^{-1}$, assuming a finite energy-dependent lifetime $\tau(\epsilon)$, we get in the quasi-particle approximation,

$$\chi_{\alpha\alpha}^R(\omega) = \sum_{k\ell ss'} |\Xi_{k\ell ss'}^\alpha|^2 \left[ \frac{f(\epsilon_{k\ell s} - \mu) - f(\epsilon_{k\ell-1s'} - \mu)}{\epsilon_{k\ell s} - \epsilon_{k\ell-1s'} + \hbar\omega + i\Gamma_{k\ell ss'}} \right]. \quad (8)$$

Here $\Gamma_{k\ell ss'} = \frac{\hbar}{2}\left[\tau(\epsilon_{k\ell s})^{-1} + \tau(\epsilon_{k,\ell-1,s'})^{-1}\right]$ is the inverse lifetime and $f$ is the Fermi–Dirac distribution. Explicit expressions for the matrix elements for intraband transitions ($s = s'$), defining $\theta_{kx}^\pm = \text{atan2}\left(x \pm \frac{1}{2}, kR\right)$ and introducing shorthand notation $x \equiv \ell - \eta$, are

$$|\Xi_{kxss}^B|^2 = \sin^2\left(\frac{\theta_{kx}^+ + \theta_{kx}^-}{2}\right),$$
$$|\Xi_{kxss}^E|^2 = \cos^2\left(\frac{\theta_{kx}^+ - \theta_{kx}^-}{2}\right). \quad (9)$$

In the presence of a finite elastic disorder $\nu$, the lifetime acquires the form

$$\frac{\hbar}{2\tau(\epsilon)} = \frac{\nu|\epsilon|}{2\pi\hbar v_F R} \sum_l \frac{\Theta\left[\epsilon^2 - \hbar^2\Omega^2\left(l + \frac{1}{2} - \eta\right)^2\right]}{\sqrt{\epsilon^2 - \hbar^2\Omega^2\left(l + \frac{1}{2} - \eta\right)^2}}, \quad (10)$$

where $\Theta(x)$ is the step function. The divergences in the expression result from van-Hove singularities associated with the bottom of the subbands ($k \to 0$), which enhance the effectiveness of the disorder scattering near these energies.

Explicit expressions for $\gamma^\alpha$ can be found using a set of approximations. As we concentrate on microwave transitions, the photon frequency is in the range $\omega \ll \Omega$. For positive chemical potential $\mu \gg \hbar\Omega$, only the positive subbands participate in the transitions, i.e., $s = s' = +$. The van-Hove singularities affect our result only if the chemical potential hits the bottom of a band ($\mu^\star = |l - \eta + 1/2|$ for some $l$, where we defined $\mu^\star = \mu/\hbar\Omega$) and simultaneously a transition becomes resonant around the chemical potential ($\ell - \eta \simeq \omega^\star \mu^\star$ with $\omega^\star = \omega/\Omega$; for a derivation of the latter see the discussion around Eq. (35)). Otherwise we can approximate $\Gamma \simeq \nu^\star \mu$, where we defined $\nu^\star = \nu/(\hbar^2 v_F^2)$, which will give us the main behavior. To make further analytical progress we write $F(k,j) \equiv (\epsilon_{k\ell s} - \epsilon_{k\ell-1 s'} - j\hbar\omega)/\hbar\Omega$, where $j \in \pm$, and change integration variables from $k$ to $F$,

$$\frac{L}{2\pi}\int_{-\infty}^{\infty} dk \to \hbar\Omega \int_{-j\omega^\star}^{|x+\frac{1}{2}|-|x-\frac{1}{2}|-j\omega^\star} dF \rho(F + j\omega^\star), \quad (11)$$

where $\rho$ is the reduced density of states for the specific transition

$$\rho(\lambda) = \frac{L}{2\pi\hbar v_F} \frac{\text{sgn}(\lambda)|\lambda^4 - 4x^2|}{\lambda^2 \sqrt{(\lambda^2 - 4x^2)(\lambda^2 - 1)}}. \quad (12)$$

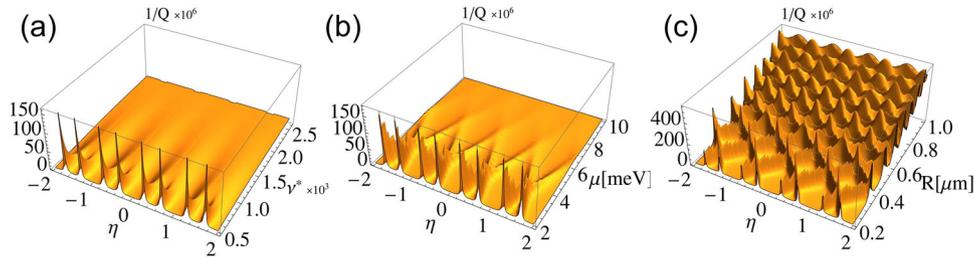

**Fig. 3 Dependence of the inverse Q-factor on the parameters of the nanowire and the external flux.** Dependence of the inverse Q-factor on the flux $\eta$ for different parameters at an anti-node of the electric field: **a** the dimensionless disorder strength $\nu^\star = \nu/\hbar^2 v_F^2$; **b** the chemical potential $\mu$; and, **c** the radius of the topological insulator nanowire (TINW) $R$. Here the length of the TINW $L = 10\,\mu$m, the Fermi velocity $v_F = 10^5$ m s$^{-1}$, the zero-point electric field $E_{zp} = 1$V m$^{-1}$, the frequency of the resonator $\omega = 0.05\Omega$ (with $\Omega = v_F/R$) and the bulk dielectric constant $\epsilon = 100$. The other parameters used are $\nu^\star = 0.0005$ (**b**, **c**), $\mu = 2$ meV (**a**, **c**) and $R = 200$ nm (**a**, **b**).

At zero temperature, the Fermi functions further constrain the range of $F$ to be between $\frac{x}{\mu^\star} - j\omega^\star \pm \frac{x^2}{2(\mu^\star)^3}$, where $\theta_{kx}^\pm \simeq \pi/2$, so Eq. (6) finally becomes $\gamma^\alpha(\omega) = \frac{2g_\alpha^2}{\hbar} \sum_{j \in \pm}[-j\text{Im}\chi_{\alpha\alpha}^R(j\omega)]$ with

$$\text{Im}\chi_{\alpha\alpha}^R(\omega) \simeq \sum_{\{x|\,|x|+\frac{1}{2}<\mu^\star\}} \rho(x/\mu^\star) \sum_{j \in \pm} j \arctan\left(\frac{\frac{x}{\mu^\star} + \omega^\star + j\frac{x^2}{2(\mu^\star)^3}}{\nu^\star\mu^\star}\right), \quad (13)$$

As it stands, Eq. (13), encapsulates the response of the TINW to the photons in the absence of electron-electron interactions. This captures the case of the response to the external oscillating magnetic field, $\gamma^B(\omega)$. The situation is somewhat analogous to the ESR effect, which enables the extraction of the gyromagnetic ratio of the electron in a material. The technique is based on measuring the energy gap generated between the two spin states of the electron under the influence of an external magnetic field, including recent advances that allow to measure transitions in nanometer structures with only few electrons[26–28]. Here, analogously, the applied flux generates a pattern of energy splittings, generating orbital magnetic resonances in the ballistic limit, two orders of magnitude larger than the bare electron spin response.

**Screening and response to electric field.** To capture correctly the response to the electric field, we need to include the Coulomb interaction between the electrons on the surface of the TINW. Once we turn electron-electron interactions on, the motion of the electrons in response to the external electric field can lead to screening of this field. This effect can be calculated using the random phase approximation (RPA), as we now turn to describe. In the RPA we replace the correlation function $\chi_{EE}^R$ by the function $\chi_{RPA}^R$

$$\chi_{RPA}^R(\omega) = \frac{\chi_{EE}^R(\omega)}{1 - W\chi_{EE}^R(\omega)}, \quad (14)$$

where $W$ is the $\Delta k \to 0$, $|\Delta \ell| = 1$ Fourier component of the Coulomb interaction on the surface of cylinder, $\frac{2e^2}{\epsilon L} I_{\Delta\ell}(|\Delta k|R) K_{\Delta\ell}(|\Delta k|R)$[29], where $I_l(x)$ and $K_l(x)$ are the modified $l$'th Bessel functions of the 1st and 2nd kind. Substituting $W = \frac{e^2}{\epsilon L}$, the excitation rate becomes,

$$\gamma_{RPA}^E(\omega) = \frac{2g_E^2}{\hbar} \sum_{j \in \pm} \frac{(-j)\text{Im}\chi_{EE}^R(j\omega)}{\left(1 - W\text{Re}\chi_{EE}^R(j\omega)\right)^2 + \left(W\text{Im}\chi_{EE}^R(j\omega)\right)^2}. \quad (15)$$

The imaginary part of $\chi_{\alpha\alpha}^R(\omega)$ is given by Eq. (13), while its real part can be written approximately as

$$\text{Re}\chi_{\alpha\alpha}^R(\omega) \simeq \sum_{\{x|\,|x|+\frac{1}{2}<\mu^\star\}} \rho(x/\mu^\star) \ln\left|\frac{(\nu^\star\mu^\star)^2 + \left(\frac{x}{\mu^\star} + \omega^\star - \frac{x^2}{2(\mu^\star)^3}\right)^2}{(\nu^\star\mu^\star)^2 + \left(\frac{x}{\mu^\star} + \omega^\star + \frac{x^2}{2(\mu^\star)^3}\right)^2}\right|. \quad (16)$$

The effectiveness of screening is dependent on the dielectric constant of the TINW. Recent studies show that the bulk dielectric constant in TIs is very high ($\epsilon \sim 100$)[30–32], which reduces the screening and, as we next discuss, can lead to measurable peaks.

## Discussion

We now discuss our results from theoretical and experimental perspectives.

The Q-factor of the resonator can be written as $Q = \omega/(\kappa + \gamma(\omega))$, where $\kappa$ is the rate of the photon loss of the resonator due to other processes (state-of-the-art resonators achieve $\omega/\kappa \sim 10^6$) and $\gamma(\omega) = \gamma^E(\omega) + \gamma^B(\omega)$ is the total absorption in the TINW per the Matthiessen rule. In a typical experiment it would be straightforward to detect dissipation when the TINW-induced dissipative response exceeds $1/Q \sim \kappa/\omega$. As can be seen in Fig. 3, the electrically induced absorption rate manifests as peaks in $1/Q$ measured as a function of the flux $\eta$. The ratio between the magnetic and electric matrix element is $g_B/g_E = v_F/c$ where $c$ is the speed of light; the magnetic resonances at the magnetic field anti-node are therefore much smaller, amounting to small changes in $1/Q$ in extremely clean samples, which could be challenging to detect.

In Fig. 3 we show how the response depends on other relevant parameters: the strength of disorder, the chemical potential, and the radius of the TINW. The disorder mostly increases the scattering and leads to a smearing of the peaks, which remain clearly visible as long as the circumference of the TINW is smaller than the scattering length $l_e = v_F\tau(\mu)$, i.e., $\nu^\star < 1/(2\pi\mu^\star)$ (Fig. 3a). Increasing the chemical potential will increase the density of states but also decrease the scattering length, and this interplay is displayed in Fig. 3b. Finally, while the dipole matrix element increases linearly with the radius, when the circumference becomes longer than the typical disorder length, the peaks will get smeared (Fig. 3c). The magnitude of the maximum of the peaks can be approximated by the following formula

$$\gamma^E(\omega) = \frac{\frac{2g_E^2}{\hbar}\frac{2L}{\pi\hbar v_F}\frac{\mu}{\hbar\omega}\arctan\left(\frac{\hbar^2\omega^2}{2\nu^*\mu^2}\right)}{\left(1+\frac{e^2}{\epsilon}\frac{4R\mu}{\pi\hbar^2 v_F^2}\right)^2 + \left(\frac{e^2}{\epsilon}\frac{2}{\pi\hbar^2 v_F}\frac{\mu}{\hbar\omega}\arctan\left(\frac{\hbar^2\omega^2}{2\nu^*\mu^2}\right)\right)^2}, \quad (17)$$

which was derived on resonance ($\ell - \eta = \omega^*\mu^*$) by taking $\text{Re}\chi_{\alpha\alpha}^R(\omega)$ in the continuum, clean limit of the two-dimensional Dirac case[33], $\text{Re}\chi_{\alpha\alpha}^R(\omega) \simeq -\frac{4LR\mu}{\pi\hbar^2 v_F^2}$. The required parameters appear to be compatible with available TINWs[9,34–38].

While we assumed a cylindrical geometry in our calculations, recent studies demonstrate that the relevant properties of TINWs are conserved for anisotropic[39] or noncylindrical[40] samples.

## Conclusions

We studied the EM response of a TINW embedded within a resonator. When the TINW is placed in an anti-node of the electric field, the surface response generates a series of peaks in the Q-factor of the resonator, owing to dipole transitions that remain unscreened due to the anomalously large dielectric constant of the bulk TI. The applied flux generates a pattern of energy splittings that results in orbital electromagnetic resonances in the ballistic limit; these broaden in the presence of disorder.

These considerations demonstrate that the resonator can probe the confinement of the two-dimensional Dirac fermions on the surface of the TINW, extracting valuable information about the excitation spectrum of the TINW, while avoiding the need to attach contacts. In conventional transport experiments, it is well known that the contacting itself may alter the properties of the TINW, while the fabrication steps required for contacting may also affect the TINW. Instead, coupling the TINW to a resonator allows a 'wireless' method of interrogating the energy spectrum of the surface states, while the bulk dielectric environment decreases the screening and supports the observation of peaks. Similarly to transport experiments, we anticipate that the effect of the properties of the topological surface states, including their spin-momentum locking and presence of Berry phase, could also be reflected in the peak structure when quantum interference effects of diffusive trajectories are taken into account. It is also of interest to extend this method to other geometries and materials.

## Methods

**Derivation of the Hamiltonian**. We start with the TI, and write the Hamiltonian governing its surface states as[22,23]

$$\tilde{\mathcal{H}}_{\text{nw}} = \frac{v_F}{2}[\hbar\nabla \cdot \mathbf{n} + \mathbf{n}\cdot(\mathbf{p}\times\boldsymbol{\sigma}) + (\mathbf{p}\times\boldsymbol{\sigma})\cdot\mathbf{n}], \quad (18)$$

where $\mathbf{p} = -i\hbar\nabla$ is the momentum operator, $\boldsymbol{\sigma} = (\sigma_x,\sigma_y,\sigma_z)$ are the Pauli operators representing the electron spin via $\tilde{\mathbf{S}} = \frac{1}{2}\hbar\boldsymbol{\sigma}$, and $\mathbf{n}$ is the normal to the surface of the TI. We choose a cylindrical geometry for the TINW. Using the transformation $\mathcal{O} = \mathcal{V}\tilde{\mathcal{O}}\mathcal{V}^\dagger$ with $\mathcal{V} = e^{-i\varphi/2}e^{i\sigma_z\varphi/2}$ and $\tilde{\mathcal{O}}$ is a general operator presented as a matrix in spin components, we transform to the local tangent spin frame around the circumference of the TINW, so the Hamiltonian acquires the form

$$\mathcal{H}_{\text{nw}} = \hbar\Omega\left[\sigma_z\left(\hat{\ell}+\frac{1}{2}-\eta\right) - \sigma_y\hat{k}R\right]. \quad (19)$$

In the tangent frame the spin is $\mathbf{S}$, while the curvature acts to introduce a Berry-type $\pi$ phase that is equivalent to shifting the boundary conditions along the $\varphi$ direction. This Hamiltonian operates on the modified wavefunction $\Psi = \mathcal{V}\tilde{\Psi}$,

$$\Psi(z,\varphi) = \frac{1}{\sqrt{2\pi L}}\sum_{k\ell s}c_{k\ell s}e^{ikz}e^{i\varphi\ell}\Psi_{k\ell s}, \quad (20)$$

where $c_{k\ell s}$ are the expansion coefficients. The associated $k,\ell$ blocks of the Hamiltonian, $\mathcal{H}_{k\ell}$, can be diagonalized using the spin rotation $U_{k\ell} = \exp\left[\frac{i}{2}\sigma_x\text{atan2}\left(\ell+\frac{1}{2}-\eta,kR\right)\right]$, according to $\mathcal{H}'_{k\ell} = U_{k\ell}\mathcal{H}_{k\ell}U^\dagger_{k\ell} = \epsilon_{k\ell}\sigma_z$ with the associated wavefunctions $\Psi'_{k\ell s} = U_{k\ell}\Psi_{k\ell s}$ written as eigenstates of $\sigma_z$, $|s\rangle$, where $s = \pm 1$ is the associated eigenvalue.

**Photon absorption rate**. Here we provide details regarding the calculations leading to the photon absorption rate by the TINW. We start with the rate of change of the number of photons in the resonator,

$$\left\langle\frac{dn_{\text{ph}}}{dt}\right\rangle = \langle\dot{a}^\dagger a + a^\dagger\dot{a}\rangle, \quad (21)$$

and employ the Heisenberg equations derived with $H = H_{\text{nw}} + H_{\text{ph}} + H_{\text{int}}$,

$$\begin{aligned}\dot{a}^\dagger &= -\frac{i}{\hbar}[a^\dagger,H]\\ &= \frac{i}{\hbar}\left(\hbar\omega a^\dagger - ig_E(\xi_E + \xi_E^\dagger) + g_B(\xi_B + \xi_B^\dagger)\right),\end{aligned} \quad (22)$$

and arrive at

$$\left\langle\frac{dn_{\text{ph}}}{dt}\right\rangle = \frac{1}{\hbar}\langle(a+a^\dagger)g_E(\xi_E + \xi_E^\dagger) + i(a-a^\dagger)g_B(\xi_B + \xi_B^\dagger)\rangle. \quad (23)$$

Adiabatically turning on the interactions, we take $\xi_\alpha(t) \to \lim_{\zeta\to 0}e^{\zeta t}\xi_\alpha(t)$ and proceeding with perturbation theory we find the following expression in the interaction picture (with respect to the resonator-TINW coupling), to second order in $\xi_{I\alpha}(t)$,

$$\begin{aligned}\frac{dn_{\text{ph}}}{dt} &= \frac{1}{\hbar}\langle\psi_I(t)|\left(a_I(t)+a_I^\dagger(t)\right)g_E(\xi_{IE}(t)+\xi_{IE}^\dagger(t)) + i\left(a_I(t)-a_I^\dagger(t)\right)g_B(\xi_{IB}(t)\\ &\quad + \xi_{IB}^\dagger(t))|\psi_I(t)\rangle = \lim_{\zeta\to 0}\frac{1}{\hbar^2}\int_{-\infty}^t dt' e^{\zeta t'}\sum_\alpha g_\alpha^2\Big\langle\Big[\Big(a_I(t)-\iota_\alpha a_I^\dagger(t)\Big)\Big(\xi_{I\alpha}(t)+\xi_{I\alpha}^\dagger(t)\Big),\\ &\quad \Big(a_I(t')+\iota_\alpha a_I^\dagger(t')\Big)\Big(\xi_{I\alpha}(t')+\xi_{I\alpha}^\dagger(t')\Big)\Big]\Big\rangle,\end{aligned} \quad (24)$$

where $|i\rangle$ is an unperturbed state of the system, $\iota_B = 1$ and $\iota_E = -1$. To calculate the commutator we insert a complete set of states $1 = \sum_j|j\rangle\langle j|$, taking into account that in the unperturbed state $|j\rangle = |j_{\text{nw}}\rangle\otimes\left|j_{\text{ph}}\right\rangle$, so we get:

$$\left\langle i_{\text{ph}}\Big|\left(a_I(t)\pm a_I^\dagger(t)\right)\Big|j_{\text{ph}}\right\rangle = \delta_{j,i+1}\sqrt{n_i+1}e^{-i\omega t} \pm \delta_{j,i-1}\sqrt{n_i}e^{i\omega t}, \quad (25)$$

for the photonic part, and in total

$$\begin{aligned}&\langle i|\Big[\left(a_I(t)-j_\alpha a_I^\dagger(t)\right)\left(\xi_{I\alpha}(t)+\xi_{I\alpha}^\dagger(t)\right),\left(a_I(t')+j_\alpha a_I^\dagger(t')\right)\left(\xi_{I\alpha}(t')+\xi_{I\alpha}^\dagger(t')\right)\Big]|i\rangle\\ &= \left((n_i+1)e^{-i\omega(t-t')} - n_i e^{i\omega(t-t')}\right)\langle i_{\text{nw}}|\xi_{I\alpha}(t)\xi_{I\alpha}^\dagger(t') + \xi_{I\alpha}^\dagger(t)\xi_{I\alpha}(t')|i_{\text{nw}}\rangle\\ &\quad + \left((n_i+1)e^{i\omega(t-t')} - n_i e^{-i\omega(t-t')}\right)\langle i_{\text{nw}}|\xi_{I\alpha}(t')\xi_{I\alpha}^\dagger(t) + \xi_{I\alpha}^\dagger(t')\xi_{I\alpha}(t)|i_{\text{nw}}\rangle.\end{aligned} \quad (26)$$

Next we need to calculate four terms of the form

$$\langle i_{\text{nw}}|\xi_{I\alpha}(t)\xi_{I\alpha}^\dagger(t')|i_{\text{nw}}\rangle = \sum_{k\ell ss'}|\Xi_{k\ell ss'}^\alpha|^2[e^{\frac{i}{\hbar}(\epsilon_{k\ell s}-\epsilon_{k\ell-1s'})(t-t')}(1-f(\epsilon_{k\ell-1s'}-\mu))f(\epsilon_{k\ell s}-\mu), \quad (27)$$

The integrations over time can now be performed, imposing energy conservation in the transitions,

$$\begin{aligned}&\lim_{\zeta\to 0}\int_{-\infty}^t dt' e^{\zeta t'}\cos\left[\left(\frac{\epsilon_{k\ell s}-\epsilon_{k\ell-1s'}}{\hbar}\pm\omega\right)(t-t')\right]\\ &= \lim_{\zeta\to 0}\hbar e^{\zeta t}\left(\frac{i}{\epsilon_{k\ell s}-\epsilon_{k\ell-1s'}\pm\hbar\omega+i\hbar\zeta} - \frac{i}{\epsilon_{k\ell s}-\epsilon_{k\ell-1s'}\pm\hbar\omega-i\hbar\zeta}\right).\end{aligned} \quad (28)$$

In the clean, noninteracting limit we get a delta function $2\pi\delta(\epsilon_{k\ell s}-\epsilon_{k\ell-1s'}\pm\hbar\omega)$. Collecting all terms, we arrive at the expression

$$\begin{aligned}\frac{dn_{\text{ph}}}{dt} &= -\frac{i}{\hbar}n_i\sum_\alpha g_\alpha^2\sum_{j\in\pm}j\left[\chi_{\alpha\alpha}^R(j\omega) - \chi_{\alpha\alpha}^{R*}(j\omega)\right]\\ &\quad + \frac{2\pi}{\hbar}\sum_\alpha g_\alpha^2\sum_{k\ell ss'}|\Xi_{k\ell ss'}^\alpha|^2\Big[\left(1-f(\epsilon_{k\ell-1s'}-\mu)\right)f(\epsilon_{k\ell s}-\mu)\delta(\epsilon_{k\ell s}-\epsilon_{k\ell-1s'}-\hbar\omega)\\ &\quad + \left(1-f(\epsilon_{k\ell s}-\mu)\right)f(\epsilon_{k\ell-1s'}-\mu)\delta(\epsilon_{k\ell s}-\epsilon_{k\ell-1s'}+\hbar\omega)\Big],\end{aligned} \quad (29)$$

where the last transition is derived in the clean, noninteracting limit. Neglecting the parts that are not proportional to $n_i$, that describe the spontaneous emission of photons from the TINW into the resonator, we finally get

$$\frac{dn_{\text{ph}}}{dt} \simeq -n_{\text{ph}}\sum_\alpha \gamma^\alpha(\omega), \quad (30)$$

where we defined the TINW excitation rate, $\gamma^\alpha(\omega)$,

$$\gamma^\alpha(\omega) = \frac{i}{\hbar}g_\alpha^2\sum_{j\in\pm}\left[j\chi_{\alpha\alpha}^R(j\omega) - \text{h.c.}\right]. \quad (31)$$

**Resolving the energy constraint**. In the clean limit, the energy constraint imposed on the transitions is

$$\epsilon_{k\ell s} - \epsilon_{k,\ell-1,s'} = \hbar\omega, \quad (32)$$

with $\epsilon_{k\ell s} = s\hbar\Omega\sqrt{\left(\ell + \frac{1}{2} - \eta\right)^2 + (kR)^2}$. To solve this constraint for the momentum, we rewrite it in the form,

$$s\sqrt{\ell_+^2 + (kR)^2} - s'\sqrt{\ell_-^2 + (kR)^2} = \omega^\star, \quad (33)$$

where $\omega^\star = \omega/\Omega$ and $\ell_\pm = \ell \pm \frac{1}{2} - \eta$. Reordering terms and squaring the equation twice, we can solve for $k$,

$$k = \pm \frac{\sqrt{[1-(\omega^\star)^2][4(\ell-\eta)^2 - (\omega^\star)^2]}}{2\omega^\star R} \equiv \pm k_{\ell,\eta}. \quad (34)$$

Next we set $k = k_{\ell,\eta}$ in the expression of the energy, to find the participating energies in terms of $\omega$

$$\epsilon_{k,\ell-\frac{1}{2}\pm\frac{1}{2},s'}\Big|_{k=k_{\ell,\eta}} = s\hbar\Omega \frac{|2(\ell-\eta) \pm (\omega^\star)^2|}{2\omega^\star}. \quad (35)$$

To find the reduced density of states we resolve the delta function imposing the energy constraint in terms of the participating momenta,

$$\delta\left(\epsilon_{k\ell s} - \epsilon_{k,\ell-1,s'} - \hbar\omega\right)$$
$$= \frac{[\delta(k-k_{\ell,\eta}) + \delta(k+k_{\ell,\eta})]\Delta_{\ell-\eta,ss'}(\omega^\star)}{|\partial_k \epsilon_{k\ell s} - \partial_k \epsilon_{k,\ell-1,s'}|}, \quad (36)$$

where $\Delta$ is given by

$$\Delta_{xss'}(\omega^\star) = \begin{cases} \delta_{s,\text{sgn}(x)\text{sgn}(\omega^\star)}\delta_{s,s'}, & |\omega^\star| < \min(1, 2|x|), \\ \delta_{s,\text{sgn}(\omega^\star)}\delta_{s,-s'}, & |\omega^\star| > \max(1, 2|x|), \\ 0 & \text{otherwise}, \end{cases} \quad (37)$$

and encodes the regions where the momentum solutions to Eq. (33) exist. Once multiplied by the density of states in $k$-space, the coefficient $\frac{L}{\pi}\left(|\partial_k \epsilon_{k\ell s} - \partial_k \epsilon_{k,\ell-1,s'}|\right)^{-1}$ becomes the reduced density of states $\rho(\omega^\star)$. In this limit the maximum of the peak is

$$\gamma^\alpha(\omega) = \frac{2L g_\alpha^2}{\hbar^2 v_F} \frac{\mu}{\hbar\omega} \tanh\left(\frac{\hbar\omega}{4k_B T}\right). \quad (38)$$

### Data availability
The authors declare that the main data supporting the findings of this study are available within the article. Extra data are available from the corresponding author upon request.

### Acknowledgements
This project has received funding from the European Unions Horizon 2020 research and innovation programme under grant agreement No. 766714. S.A.H. and E.Gr. acknowledge support from the Israel Science Foundation under grant 1626/16.

### Author contributions

S.A.H., E.Gi., S.E.G. and E.Gr. contributed to all aspects of the research; S.A.H. performed all the calculations.

### Competing interests

The authors declare no competing interests.

### Additional information

**Correspondence** and requests for materials should be addressed to Eytan Grosfeld.

**Peer review information** *Communications Physics* thanks Kristof Moors for their contribution to the peer review of this work.

**Reprints and permission information** is available at http://www.nature.com/reprints

**Publisher's note** Springer Nature remains neutral with regard to jurisdictional claims in published maps and institutional affiliations.

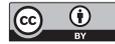